\documentclass[aps,prl,twocolumn,showpacs]{revtex4}

\usepackage{graphicx, amsmath}

\begin{document}

\bibliographystyle{apsrev}

\title{Entanglement between the future and past in the quantum vacuum}

\author{S. Jay Olson}
 \email{j.olson@physics.uq.edu.au}
 \author{Timothy C. Ralph}
\affiliation{Centre for Quantum Computing Technology, Department of Physics, University of Queensland, St Lucia, Queensland 4072, Australia}

\date{\today}

\begin{abstract}
We note that massless fields within the future and past light cone may be quantized as independent systems.  We show that the vacuum is an entangled state of these systems, exactly mirroring the known entanglement between the spacelike separated Rindler wedges.   We describe a detector which exhibits a thermal response to the vacuum when switched on at $t=0$.  The feasibility of experimentally detecting this effect is discussed.

\end{abstract}

\maketitle

A basic and far-reaching property of the quantum vacuum is that it is an entangled state \---- a fact underlying an impressive number of theoretical insights and predictions~\cite{birrell1}.  In the case of flat Minkowski spacetime, this is typically shown in the context of the Unruh effect~\cite{unruh1, crispino1}.  There, the vacuum state of the field can be written as an entangled state between two sets of modes, respectively spanning two space-time wedges, known as the Rindler wedges (see Fig. 1).  A uniformly accelerated observer sees only one set of Rindler modes.  The tracing out of the unobserved Rindler modes leads to the prediction that such an accelerated observer sees a thermalized vacuum.  Having been predicted over 30 years ago, the Unruh effect remains unobserved.

Here, our main result is to demonstrate that precisely the same entanglement exists between the fields within the future and past light cone (F and P, respectively) as between the left and right Rindler wedges (L and R), and that the Unruh effect can be mapped onto an equivalent vacuum thermal effect for an inertial observer constrained to interact with the field in only the future or the past.  We will demonstrate this result for a scalar field in 2-d spacetime, though the result is general for massless fields.  Dimensional analysis suggests that observation of this effect may be within range of current technology.

This paper is organized as follows:  We first note that massless fields in F and P may be quantized as independent systems, and then describe our coordinatization of spacetime, and the mode functions living in each quadrant.  We then express the state of the Minkowski vacuum restricted to F and P in terms of these modes, and note entanglement.  An Unruh-DeWitt detector is then described, which shows a thermal response to these modes in F (or P), and a related prediction of Martinetti and Rovelli~\cite{rovelli10} is interpreted in light of this result.  The feasibility of an experimental observation of this effect is discussed, based on dimensional analysis.  We then offer some conclusions.

\begin{figure}
 \centering
 \includegraphics[scale=0.25]{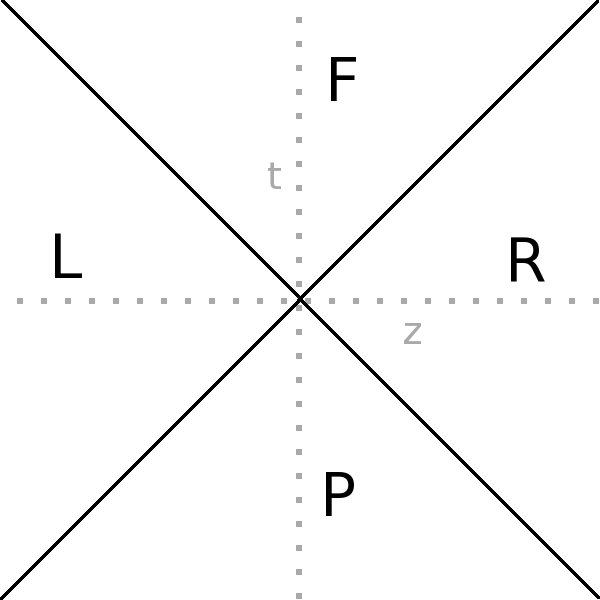}
 \caption{Spacetime divided into four quadrants, consisting of regions contained by the future and past light cones (F and P), and the right and left Rindler wedges (R and L).}
 \label{fig:  Quadrants of Minkowski spacetime}
\end{figure}

\textbf{Past/future commutator and independent systems:}  The concept of entanglement between the left and right Rindler wedges rests on the fact that the fields within may be quantized as independent systems.  This is expressed through the vanishing of the Pauli-Jordan function, $i \Delta (x - y) = [\hat{\phi}(x) , \hat{\phi}^{\dagger}(y)]$ for spacelike intervals.  This general feature holds for both massive and massless fields.

In the case of massless fields, however, the Pauli-Jordan function $\Delta (x - y)$ vanishes for all but lightlike intervals, $(x - y)^2 = 0$~\cite{heitler1}.  In particular, it vanishes for timelike intervals.  This will allow us to employ known quantization techniques in both F and P, and to regard the fields there as independent systems.

\textbf{Coordinates:} We now break spacetime into four quadrants, F P R and L, and introduce four coordinate systems valid in each quadrant.  These will then be related to the light cone coordinates which are valid over all spacetime.

In F, we adopt the following coordinates $\eta$ and $\zeta$:
\begin{eqnarray*}
 t  &=&  a^{-1} e^{a \eta} \cosh(a \zeta) \\
 z  &=&  a^{-1} e^{a \eta} \sinh(a \zeta) .
\end{eqnarray*}

In P, we will use the coordinates $\bar{\eta}$ and $\bar{\zeta}$:
\begin{eqnarray*}
 t = - a^{-1} e^{a \bar{\eta}} \cosh(a \bar{\zeta}) \\
 z = - a^{-1} e^{a \bar{\eta}} \sinh( a \bar{\zeta}).
\end{eqnarray*}

These are to be compared with the usual Rindler coordinates $\tau$ and $\epsilon$, in R:
\begin{eqnarray*}
t  =  a^{-1} e^{a \epsilon} \sinh(a \tau) \\
z  =  a^{-1} e^{a \epsilon} \cosh(a \tau).
\end{eqnarray*}

As well as $\bar{\tau}$ and $\bar{\epsilon}$, in L:
\begin{eqnarray*}
t  =  -a^{-1} e^{a \bar{\epsilon}} \sinh(a \bar{\tau}) \\
z  =  -a^{-1} e^{a \bar{\epsilon}} \cosh(a \bar{\tau}).
\end{eqnarray*}

In each of these coordinate systems the metric is conformal to the Minkowski metric, and owing to the conformal invariance of the massless wave equation, the same equation holds separately in the four coordinate systems and their respective quadrants, namely:
\begin{eqnarray*}
\left( \frac{\partial^2}{\partial \eta^2} - \frac{\partial^2}{\partial \zeta^2}\right)_{F} \phi = 0, && \left( \frac{\partial^2}{\partial \tau^2} - \frac{\partial^2}{\partial \epsilon^2}\right)_{R} \phi = 0 \\
\left( \frac{\partial^2}{\partial \bar{\eta}^2} - \frac{\partial^2}{\partial \bar{\zeta}^2}\right)_{P} \phi = 0, && \left( \frac{\partial^2}{\partial \bar{\tau}^2} - \frac{\partial^2}{\partial \bar{\epsilon}^2}\right)_{L} \phi = 0.
\end{eqnarray*}

We now introduce the light-cone coordinates, valid throughout all spacetime:
\begin{eqnarray*}
V = t + z, & U = t - z
\end{eqnarray*}

and their analogs in the above four coordinate systems:
\begin{eqnarray*}
(F) & \nu = \eta + \zeta, & \mu = \eta - \zeta \\
(P) & \bar{\nu} = -\bar{\eta} - \bar{\zeta}, & \bar{\mu} = - \bar{\eta} + \bar{\zeta} \\
(R) & \chi = \tau + \epsilon, & \kappa = \tau - \epsilon \\
(L) & \bar{\chi} = - \bar{\tau} - \bar{\epsilon}, & \bar{\kappa} = - \bar{\tau} + \bar{\epsilon}.
\end{eqnarray*}

These are related in the following way:
\begin{eqnarray*}
(F) & V = a^{-1} e^{a \nu}, & U = a^{-1} e^{a \mu} \\
(P) & V = - a^{-1} e^{- \bar{\nu}}, & U = - a^{-1} e^{-a \bar{\mu}} \\
(R) & V = a^{-1} e^{a \chi}, & U = -a^{-1} e^{-a \kappa} \\
(L) & V = - a^{-1} e^{- a \bar{\chi}}, & U = a^{-1} e^{a \bar{\kappa}}.
\end{eqnarray*}

\textbf{Field expansion, Bogoliubov transformations, and entanglement:}  Using the light-cone coordinates, the field may be expanded in plane waves:
\begin{eqnarray*}
\hat{\Phi}(V, U) &=& \int_{0}^{\infty} \frac{dk}{ (4 \pi k)^{1/2}}  [ \hat{a}^{1}_{k} e^{-ikV} + \hat{a}^{1 \dagger}_{k} e^{ikV} \\
 &+& \hat{a}^{2}_{k} e^{-ikU} + \hat{a}^{2 \dagger}_{k} e^{ikU}  ].
\end{eqnarray*}

Since all $\hat{a}^{1}$'s commute with all $\hat{a}^{2}$'s (corresponding to left and right moving modes, respectively), we make a common simplification, and treat only the left moving sector of the field, $\hat{\Phi}(V) = \int_{0}^{\infty} \frac{dk}{ (4 \pi k)^{1/2}}  [ \hat{a}^{1}_{k} e^{-ikV} + \hat{a}^{1 \dagger}_{k} e^{ikV}]$, with the understanding that analogous results hold for the right moving sector $\hat{\Phi}(U)$ as well.

We can similarly expand $\hat{\Phi}(V)$ in terms of the following sets of functions, in their respective quadrants.  The Rindler modes:
\begin{eqnarray*}
g_{\omega}^{R}(\chi) = (4 \pi \omega)^{-1/2} e^{-i \omega  \chi} \\
g_{\omega}^{L}(\bar{\chi}) = (4 \pi \omega)^{-1/2} e^{-i \omega \bar{\chi}}
\end{eqnarray*}

and their analogs in the future and past:
\begin{eqnarray*}
g_{\omega}^{F}(\nu) = (4 \pi \omega)^{-1/2} e^{-i \omega \nu} \\
g_{\omega}^{P}(\bar{\nu}) = (4 \pi \omega)^{-1/2} e^{-i \omega \bar{\nu}}.
\end{eqnarray*}

We now note that $g_{\omega}^{F}(\nu)$ is in fact the same solution as $g_{\omega}^{R}(\chi)$, extended from R into F \---- a fact pointed out by Gerlach~\cite{gerlach1, crispino1} in the more general context of massive fields.  This can be seen by expanding these functions in terms of plane waves:
\begin{eqnarray}
 \theta(V) g_{\omega}^{R}(\chi) = \int_{0}^{\infty} \frac{dk}{(4 \pi k)^{1/2}}\left( \alpha^{R}_{\omega k} e^{-ikV} + \beta^{R}_{\omega k} e^{ikV} \right) \\
 \theta(V) g_{\omega}^{F}(\nu) = \int_{0}^{\infty} \frac{dk}{(4 \pi k)^{1/2}}\left( \alpha^{F}_{\omega k} e^{-ikV} + \beta^{F}_{\omega k} e^{ikV} \right) .
\end{eqnarray}

Now note that $g_{\omega}^{R}(\chi)$ and $g_{\omega}^{F}(\nu)$ are identical as functions of $V$, since $\chi(V) = \nu(V)$, and so they are built up of exactly the same plane waves.  The same relationship holds for $g_{\omega}^{L}(\chi)$ and $g_{\omega}^{P}(\nu)$.  In other words, the Bogoliubov coefficients satisfy:
\begin{eqnarray*}
 \alpha^{F}_{\omega k} = \alpha^{R}_{\omega k}, && \beta^{F}_{\omega k} = \beta^{R}_{\omega k} \\
 \alpha^{P}_{\omega k} = \alpha^{L}_{\omega k}, && \beta^{P}_{\omega k} = \beta^{L}_{\omega k}\end{eqnarray*}

Thus, the well-known relations between Bogoliubov coefficients in R and L are duplicated in F and P.  In particular, solving 1-2 (and the analogous P and L relations) leads to Bogoliubov coefficients which satisfy the relations,  $\beta^{P}_{\omega k} = - e^{-\pi \omega / a} \alpha^{F \ast}_{\omega k}$ and $\beta^{F}_{\omega k} = - e^{-\pi \omega / a} \alpha^{P \ast}_{\omega k}$.

From this point forward, the demonstration of future/past entanglement of the Minkowski vacuum is \textit{exactly the same} as the standard demonstration of right/left entanglement, with a change of labels $R \rightarrow F$ and $L \rightarrow P$.  We review the basic argument, and refer the reader to Crispino, Higuchi, and Matsas~\cite{crispino1} for detail.

Using the above Bogoliubov relations, the following pure-positive frequency modes can be defined:
\begin{eqnarray*}
G_{\omega}(V) = \theta(V)g^{F}_{\omega}(\nu) + \theta(-V) e^{- \pi \omega / a} g^{P \ast}_{\omega}(\bar{\nu}) \\
\bar{G}_{\omega}(V) = \theta(-V)g^{P}_{\omega}(\bar{\nu}) + \theta(V) e^{- \pi \omega / a} g^{F \ast}_{\omega}(\nu).
\end{eqnarray*}

When the field is expanded in terms of these functions, the annihilation operators for the $G$ and $\bar{G}$ quanta can readily be seen to be $(\hat{a}^{F}_{\omega} - e^{- \pi \omega /a} \hat{a}^{P \dagger}_{\omega})$ and $(\hat{a}^{P}_{\omega} - e^{- \pi \omega / a} \hat{a}^{F \dagger}_{\omega} )$ (where the $\hat{a}$'s here refer to the $g$-quanta).  Since both $G$ and $\bar{G}$ are pure positive frequency functions of Minkowski time, their vacuum coincides with the Minkowski vacuum $|0_{M} \rangle$, and we obtain the relations:
\begin{eqnarray}
(\hat{a}^{F}_{\omega} - e^{- \pi \omega /a} \hat{a}^{P \dagger}_{\omega}) |0_{M} \rangle = 0 \\
(\hat{a}^{P}_{\omega} - e^{- \pi \omega / a} \hat{a}^{F \dagger}_{\omega} ) |0_{M} \rangle = 0 ,
\end{eqnarray}
which in turn imply the following:
\begin{eqnarray}
(\hat{a}^{F \dagger}_{\omega} \hat{a}^{F}_{\omega} - \hat{a}^{P \dagger}_{\omega} \hat{a}^{P}_{\omega}) |0_{M} \rangle = 0.
\end{eqnarray}

Define the vacuum $|0_{T} \rangle$ (which is the Rindler vacuum, but defined for our purposes within the future and past light cone, F-P) to satisfy $\hat{a}^{F}_{\omega} |0_{T} \rangle = \hat{a}^{P}_{\omega} |0_{T} \rangle = 0$.  Using the approximation that there are a discrete set of modes labeled by $\omega_{i}$, the relations 3-5 imply that the Minkowski vacuum restricted to F-P may then be expressible in the following form:
\begin{eqnarray}
|0_{M} \rangle = \prod_{i} C_{i} \sum_{n_{i}=0}^{\infty} \frac{e^{- \pi n_{i} \omega_{i} /a}}{n_{i} !}( \hat{a}^{F \dagger}_{\omega_{i}}\hat{a}^{P \dagger}_{\omega_{i}})^{n_{i}} |0_{T} \rangle,
\end{eqnarray}
which is clearly entangled.  Also mirroring the Rindler case, the state of the future (or the past) alone is a ``thermal'' state of the $g_{\omega}$-modes, where $\omega$ is a frequency with respect to the conformal time coordinate $\eta$.
\begin{eqnarray}
\hat{\rho}_{F} = \prod_{i} \left[ C_{i}^{2} \sum_{n_{i} =0}^{\infty} e^{- 2 \pi n_{i} \omega_{i} /a} |n_{i}^{F} \rangle \langle n_{i}^{F} | \right],
\end{eqnarray}
where $C_{i} = \sqrt{1 - e^{-2 \pi \omega_{i} /a}}$.

\textbf{Detectors:} The above result suggests that, in analogy with the Unruh effect, an inertial detector switched on at $t=0$ and sensitive to frequency $E$ with respect to conformal time $\eta$ should register a thermal response.  This can indeed be seen to be the case, by similar reasoning.

The Schr\"odinger equation in the conformal time $\eta$ along a motionless trajectory takes the following form:  $i \frac{\partial}{\partial \eta} \Psi = \frac{\partial t}{\partial \eta} H \Psi = atH \Psi $.  To use perturbation theory in $\eta$, we take the Hamiltonian to be $H = H_{0}/at + H_{I}$, where $H_{I}$ is the standard interaction term for an Unruh-DeWitt detector, namely $\alpha \hat{m} \hat{\phi}$.  The time scaling of the $H_{0}$ term means the detector will have a constant $\eta$-frequency gap, $E$.  Converting to $\eta$, the Schr\"odinger equation reads:
\begin{eqnarray*}
i \frac{\partial}{\partial \eta} \Psi = \left( H_{0} + e^{a \eta} H_{I}\right) \Psi
\end{eqnarray*}

The $e^{a \eta}$ factor in the interaction term means that perturbation theory will eventually break down, as the energy levels of the detector are moved closer together with time \---- eventually the coupling to the field will dominate this small energy gap.  Here, we will assume that the coupling constant $\alpha$ and choice of parameter $a$ have been made small enough so that perturbation theory remains valid over times that are large compared to any other relevant timescale appearing in the problem, and interpret infinite integrals over $\eta$ as integrals to ``arbitrarily large $\eta$ within this approximation.''

In the Heisenberg picture in $\eta$, the detector's monopole moment operator evolves like $\hat{m}(\eta) = e^{i H_{0} \eta} \hat{m} e^{-i H_{0} \eta}$, while the field operator transforms as a scalar under the change $t \rightarrow \eta$, so that $\hat{\phi}(\eta) = \hat{\phi}(t(\eta))$.  The detector response function can thus be calculated in the standard way, giving:
\begin{eqnarray*}
F(E) = \int_{-\infty}^{\infty} d \eta \int_{- \infty}^{\infty} d \eta' e^{-iE(\eta - \eta')} e^{a(\eta + \eta')} D^{+}(\eta, \eta'), 
\end{eqnarray*}
where $D^{+}(\eta, \eta')= \langle 0_{M} | \phi(\eta) \phi(\eta') |0_{M} \rangle.$

This differs from the usual form only in the presence of the $e^{a(\eta + \eta')}$ factor in the integrand, and the fact that the limits of integration correspond to a detector turned on at $t=0$.  The usual regularized form of $D^{+}(x, x') = \langle 0_{M} | \phi(x) \phi(x') |0_{M} \rangle$ is given by~\cite{birrell1}:
\begin{eqnarray*}
 D^{+}(x, x') = - \frac{1}{4 \pi^2} \left[ (t - t' - i \epsilon)^{2} - (\vec{x} - \vec{x}')^{2} \right]^{-1}.
\end{eqnarray*}

We now note that the functional form of $e^{a(\eta + \eta')}D^{+}(\eta, \eta')$ for two points on the inertial trajectory $t = a^{-1}e^{a \eta}$, $z=0$ is identical to that of $D^{+}(\tau, \tau')$ for two points on the accelerated trajectory $t = a^{-1} \sinh(a \tau )$, $z = a^{-1} \cosh(a \tau)$, up to a re-scaling of $\epsilon$.  This can be seen through a straightforward coordinate substitution, noting in particular for the inertial case that:
\begin{eqnarray*}
\frac{1}{(t - t')^{2}} &=& \frac{a^2 e^{-a(\eta + \eta')}}{e^{a (\eta - \eta')} + e^{- a (\eta - \eta')} - 2} \\
&=& \frac{a^2 e^{-a(\eta + \eta')}}{4 \sinh^{2}(\frac{a}{2}(\eta - \eta'))},
\end{eqnarray*}
and for the accelerated case:
\begin{gather*}
\frac{1}{(t - t')^{2} - (z - z')^{2}} \\
 = \frac{a^2}{(\sinh(a \tau) - \sinh(a \tau'))^{2} - (\cosh(a \tau) - \cosh(a \tau'))^{2}} \\
 = \frac{a^2}{4 \sinh^{2}(\frac{a}{2}(\tau - \tau'))}.
\end{gather*}

By correspondence to the Unruh effect, this leads to a thermal response at ``temperature'' $T = \frac{\hbar a}{2 \pi k}$ (note the lack of ``c'' compared to the Unruh case \---- more on this below).  We thus conclude that an inertial, two-state Unruh-DeWitt detector, whose energy gap is continuously scaled as $\frac{1}{at}$ responds to the Minkowski vacuum in a manner identical to an accelerated detector with fixed proper-energy gap, as their response functions are identical.

Our results have previously been hinted at in the literature by Bunch, Christensen, and Fulling~\cite{bunch1, birrell1}, who studied quantum field theory in the Milne universe (a cosmology consisting of only the future light cone) where a ``thermal correspondence'' was noted between two differently defined (but pure) vacua.  More recently, Martinetti and Rovelli~\cite{rovelli10} have predicted that a time-dependent temperature, $T = \frac{1}{2 \pi t}$, should be encountered by an observer ``born at $t=0$,'' but no interpretation was given in terms of detectors.  Recently, the response of an inertial Unruh-Dewitt detector (without frequency scaling) switched on abruptly at $t=0$ was computed by Louko and Satz~\cite{louko1}, using techniques developed by Schlicht~\cite{schlicht1} \---- the response is not obviously thermal (and not identified as such by Louko and Satz).  However, if a detector is designed to compensate for the changing temperature with a changing energy gap, then $\frac{E}{T}$ would become constant in time, and one should expect to see an effectively constant-temperature thermal response \---- exactly what we have shown here to be the case.  We believe this is the natural interpretation of Martinetti and Rovelli's result, in terms of Unruh-DeWitt detectors.  Further, our result reveals the source of thermalization via the F-P entanglement.

\textbf{Feasibility of experimental detection:}  The original Unruh effect is notoriously difficult to observe, since the temperature is so tiny for commonly accessible values of the acceleration, $a$, namely $T_{U} = \frac{\hbar a}{2 \pi c k}$.  To encounter a temperature of 1 degree Kelvin requires an acceleration on the order of $10^{20} \frac{m}{s^{2}}$.

However, we have seen that scaling the energy level of an inertial detector allows interaction with precisely the same field modes in the same thermal state.  In our case, $a$ is simply a scaling constant with units of $\frac{1}{sec.}$.  The factor of $c^{-1}$ thus disappears from the temperature, $T = \frac{\hbar a}{2 \pi k}$.  To encounter a temperature of 1 degree Kelvin thus requires $a$ on the scale of 100 gigahertz.

We also require that the energy gap of the detector is scaled over a long enough period to allow thermalization.  We imagine a detector which is scaled between times $\eta_{1}$ and $\eta_{2}$, and we demand many oscillations at the constant $\eta$-frequency $E$ of the detector, within the interaction time period.  This requirement reads:
\begin{eqnarray*}
\eta_{2} - \eta_{1} \gg E^{-1}.
\end{eqnarray*}

Expressed in ordinary $t$-times and frequencies (in which $\bar{E}_{1}$ is the initial $t$-frequency gap of the detector at $t_{1}$, etc.), this reads:
\begin{eqnarray*}
\frac{t_{2}}{t_{1}} \gg e^{\frac{1}{t_{1} \bar{E}_{1}}} = e^{\frac{1}{t_{2} \bar{E}_{2}}}.
\end{eqnarray*}

Now, if $t_{1}$ is taken to be the characteristic timescale $1/\bar{E}_{1}$, then thermalization requires $t_{2}  \gg 2.7 t_{1}$.

In other words, the most basic dimensional analysis suggests that the effect could be visible on frequency scales in the vicinity of $100$ gigahertz, scaled over a single order of magnitude.

\textbf{Interpretation and conclusions:} In contrast to earlier work~\cite{vedral1}, we view entanglement across time in a standard way \---- a property of the state of a multicomponent system.  In addition to implying measurement correlations, we also inherit some ``inverted'' causality considerations.  For example, experiments performed in P and F may act at the same location in space, but the fields are causally independent at different times.  The no-signaling theorem~\cite{ghirardi1} thus directly prevents us from exploiting F-P entanglement to make a time machine.

Although the field is causally disconnected between F and P, measurements in P (for example, projections onto $g_{\omega}$-particle number) can collapse the state of the field in F (and vice versa).  While all field disturbances propagate away from the source of the interaction at the speed of light, a kind of ``entanglement residue'' remains local, and allows for the prediction of the outcomes of future measurements at the same spatial location.

It was noted above that the entangled modes in F-P are the same mode solutions as the entangled Rindler modes in R-L.  In other words, F-P entanglement is not merely \emph{analogous} to R-L entanglement \---- it is \emph{precisely the same entanglement}, viewed in a different region of spacetime.  Recently, theoretical work has been focused on manipulating vacuum entanglement, by placing various devices in accelerated motion to interact with these modes~\cite{alsing1, bradler1, reznik1, mann1, lin1, han1} \---- these are used to illustrate ``exotic effects'' which are in principle allowed by relativistic quantum field theory.  We speculate here that due to the dimensional improvement and the lack of need for continuous acceleration, some of these effects may in fact become experimentally accessible, when converted to the equivalent interactions with the same modes in F-P.

We would like to thank Tony Downes, Achim Kempf, and Nicolas Menicucci for stimulating discussions.  In addition, we would like to thank the Defence Science and Technology Organisation (DSTO) for their support.

\bibliography{ref1}{}
\bibliographystyle{unsrt}
 
\end{document}